\documentstyle[12pt,aasms4]{article}
\received{}
\revised{}
\accepted{}
\righthead{Quantitative Morphology of UV Galaxies}
\lefthead{Kuchinski et al.}

\begin{document}

\title{Quantitative Morphology of Galaxies Observed in the Ultraviolet}

\author{L.E. Kuchinski\altaffilmark{1}, Barry F. Madore\altaffilmark{2,3}, W. L. Freedman\altaffilmark{2}, M. Trewhella\altaffilmark{1}}
\altaffiltext{1}{Infrared Processing and Analysis Center, Caltech/JPL, Pasadena, CA 91125}
\altaffiltext{2}{Observatories of the Carnegie Institute of Washington, Pasadena, CA 91101}
\altaffiltext{3}{NASA/IPAC Extragalactic Database, Caltech, Pasadena, CA 91125}

\begin{abstract}
We  present a  quantitative study  of the  far--ultraviolet  (FUV) and
optical   morphology  in   32   nearby  galaxies   and  estimate   the
``morphological  $k$--correction''  expected  if  these  objects  were
observed  unevolved at  high redshift.   Using the  common  indices of
central concentration ($C$) and rotational asymmetry ($A$) to quantify
morphology, we consider independently  two phenomena that give rise to
this  $k$--correction.   Bandshifting,  the  decrease  in  rest--frame
wavelength of  light observed through optical filters,  is explored by
measuring these indices  in several passbands for each  galaxy, and it
is  found to be  the primary  driver of  changes in  $C$ and  $A$.  In
general, the optical trend found for decreasing $C$ and increasing $A$
when going  to shorter wavelengths  extends to the FUV.   However, the
patchy nature of recent  star--formation in late--type galaxies, which
is accentuated in the FUV, results in poor quantitative correspondence
between morphologies determined in the optical and FUV.

\par We then artificially redshift our FUV images into the Hubble Deep
Field (HDF) filters to  simulate various cosmological distance effects
such  as surface brightness  dimming and  loss of  spatial resolution.
Hubble  types  of  many  galaxies   in  our  sample  are  not  readily
identifiable  at  redshifts  beyond  $z  \sim  1$,  and  the  galaxies
themselves are  difficult to detect  beyond $z \sim 3$.   Because only
features  of   the  highest  surface  brightness   remain  visible  at
cosmological distances,  the change in  $C$ and $A$  between simulated
high--$z$  galaxies  and their  unredshifted  counterparts depends  on
whether their  irregular features are primarily bright  or faint.  Our
simulations suggest that $k$--corrections  alone are indeed capable of
producing  the peculiar  morphologies observed  at high  redshift; for
example,  several  spirals  have   $C$  and  $A$  indices  typical  of
irregular/peculiar  HDF objects  viewed  at $z  \geq  2$.  We  briefly
discuss some elements  of a scheme to classify  rest--frame UV images,
mergers, protogalaxies,  and other objects for  which classical Hubble
types do not adequately encompass the existing morphology.
\end{abstract}


\section{Introduction}

\par  The morphology  of  galaxies in  local  and distant  populations
provides clues about the physical processes that shaped these systems,
either at the  time of their formation or  during their evolution over
the age of  the universe.  In the framework  of the traditional Hubble
scheme, morphology  has been correlated  with a number  of fundamental
underlying  physical  properties   (for  a  comprehensive  review  see
\cite{rob94}  and   references  therein).   Elliptical   galaxies  are
dynamically hot systems supported by their velocity dispersions, while
spiral (disk) galaxies are dynamically cold and rotationally supported
({\it  e.g.}   \cite{kor82}).   Bulge--to--disk  ratios   reflect  the
relative  importance of  dynamically  hot and  cold populations  ({\it
e.g.}  \cite{ken86}).   The  presence  of  a  bar  may  imply  secular
evolution   that   builds  spiral   bulges   or  produces   starbursts
(\cite{pfe90}; \cite{cdb96}).  Galaxies with early types in the Hubble
sequence tend to be more massive, more luminous, and have less gas and
a  lower  present--day star  formation  rate  than late--type  systems
(Roberts \& Haynes).   However, it is expected that  the morphology of
high--redshift galaxies  could be  quite different from  that observed
locally.  Theoretical scenarios of structure formation in the universe
predict a  significant fraction of mergers and  interacting systems at
high   redshift  compared   to  the   local  population   ({\it  e.g.}
\cite{bcf96}).  By  carefully comparing samples of  nearby and distant
objects,  it  may  be  possible  to identify  the  characteristics  of
protogalaxies and young star--forming systems and understand the roles
of  monolithic collapse  and mergers  in shaping  the galaxies  of the
current epoch.

\par  There are  two major  drawbacks to  using the  Hubble  scheme in
studies of galaxy evolution  from deep surveys: its subjective nature,
and  its  lack  of   descriptive  and  discriminative  power  for  the
irregular,  starbursting, and  interacting  systems that  may be  more
prevalent at high redshift.  Classification into Hubble types is based
on qualitative analysis of observable  features in each galaxy, and it
can thus differ  from one observer to another.   Multiple features are
considered  and weighted  together,  making it  difficult to  automate
classification  on  the  Hubble  sequence  and  occasionally  yielding
internally   contradictory  suggestions  of   the  type   ({\it  e.g.}
bulge--to--disk ratio  {\it vs.}  winding of the  spiral arms).   In a
comparison  of classifications of  $\sim $800  galaxies by  six expert
morphologists, \cite{nai95}  find a  dispersion in revised  Hubble $T$
index of $\sigma  \sim $1.8, where a change of  1.0 corresponds to the
difference between, {\it e.g.} Sa and Sab.  For peculiar galaxies, the
level  of  agreement  between  different experts'  classifications  is
substantially lower  (\cite{nai97a}), and  the Hubble scheme  does not
divide  the  peculiar  systems  into  any  further  categories  beyond
division  of  the faint  irregulars  with no  bulge  or  arms and  the
peculiar  galaxies that  may have  tidal features,  mergers,  or other
obvious disruptions.   This ambiguity  at the late  end of  the Hubble
sequence is particularly problematic  for surveys that probe evolution
of  the galaxy  population because  there is  evidence of  an apparent
increase  in  the  fraction  of irregular/peculiar  galaxies  at  high
redshift ({\it e.g.}  \cite{bri98}; \cite{dri98}; \cite{abr96a}).

\par  In   recent  years,  renewed   attention  has  focused   on  the
quantitative  classification of  galaxies as  an  objective, automated
measure of  their properties and evolution. The  emerging methods are,
by design,  easily applicable to  the current generation of  large CCD
imaging surveys.   Abraham et al.  (1994) developed a  numerical index
for the central concentration  ($C$) of galaxies, following the Yerkes
classification  system of  \cite{mor58} and  the  correlations between
concentration   and  Hubble   type  presented   by   \cite{okw84}  and
\cite{dfo93}.    These   authors   later   (\cite{abr96b})   added   a
quantitative  measure of galaxy  asymmetry ($A$)  to form  a two--part
classification  system  that distinguishes  between  three bins:  E/S0
galaxies,  spirals,  and  irregular/peculiar  systems.  If  data  from
different wavelengths are available,  it is possible to further divide
the latest type  bin between the Hubble Irr types  and mergers using a
correlation  between  color  and  asymmetry  that is  valid  only  for
non--interacting galaxies  (\cite{con00}; \cite{con97}).  An important
step in the development of this scheme was the classification of local
galaxies   based  on   their  optical   images,  which   provides  the
``calibration''  of  $C$  and  $A$  to  Hubble  Types  (\cite{abr96b};
\cite{bri98};  Conselice et al.  1999).  This  $A$--$C$ classification
system has been used to  study the distribution of morphologies in the
{\it   Hubble   Space    Telescope}   ($HST$)   Medium   Deep   Survey
(\cite{abr96b}),  the  Hubble  Deep  Field (HDF,  \cite{abr96a}),  the
ground--based       Canada--France      Redshift       Survey      and
Autofib/Low--Dispersion    Survey    Spectrograph   Redshift    Survey
(\cite{bri98}), and $HST$ NICMOS deep images (\cite{tep98}).  Division
of the morphological data into redshift bins demonstrates the increase
in   apparently   irregular/peculiar   galaxies   at   high   redshift
(\cite{bri98}).

\par  To study  galaxy evolution,  it is  important to  understand the
behavior  of $C$  and $A$  as a  function of  wavelength in  the local
population  before  applying these  indices  to  a  sample of  objects
observed  at a  range of  redshifts ({\it  i.e.}  different look--back
times).   At long  rest wavelengths,  where  stellar light  is a  good
tracer  of mass,  low degrees  of central  concentration  and symmetry
reflect a lack of  dynamical organization that may characterize either
interacting systems or those in  the process of formation.  At shorter
wavelengths,  asymmetry and  patchiness are  more likely  to highlight
cases in  which dust and/or  recent localized star  formation strongly
influence  the observed morphology.   Thus it  is not  surprising that
$B$--band  images already  show lower  $C$ and  higher $A$  than their
$R$--band counterparts (\cite{bri98};  \cite{con97}).  In light of the
physically different  regimes probed  by optical and  UV light,  it is
necessary to understand how this morphological trend extends to the UV
to determine if simple  quantitative $k$--corrections are feasible for
the concentration and asymmetry indices.

\par Because the rest--frame  far--ultraviolet (FUV) light of galaxies
is   redshifted  into   optical  filters   at  $z\sim   3$   and  into
near--infrared (NIR)  ones at $z\sim  10$, the UV morphology  of local
galaxies  provides  an  essential  basis for  interpreting  images  of
high--redshift   systems.    However,   this  wavelength   regime   is
inaccessible from the ground, and the availability of UV galaxy images
obtained from rocket and orbital missions has been limited until quite
recently  ({\it  e.g.}   \cite{oco97}  and references  therein).   The
existing data  suggest that  UV and optical  morphology are  not often
well--coupled ({\it  e.g.}  \cite{oco97};  Marcum et al.   1997, 2001;
\cite{kuc99}, hereafter K00), and the interested reader is referred to
an early demonstration  of these facts by Bohlin  et al. (1991).  Thus
it is difficult to interpret the influence of bandshifting on observed
differences between the optical characteristics of nearby galaxies and
the recorded  rest--frame UV appearance of distant  ones.  Compared to
their  optical classifications,  spiral galaxies  generally  appear to
have later  Hubble types in  the UV (\cite{oco97}; K00).   Spiral arms
and star--forming rings  are more prominent in the  UV than at optical
wavelengths, while bulges and bars are much fainter, nearly invisible,
in  the  UV (\cite{wal97};  K00).   The  UV  radial profiles  of  disk
galaxies  are   flatter  than   optical  profiles,  and   the  central
concentration   appears  to   be  lower   in  the   UV  (\cite{oco97};
\cite{fan97a}; K00).   FUV images of late--type  galaxies often appear
much  more  fragmented than  the  optical  view (\cite{oco97}),  which
raises the possibility of  mistaking an ordinary irregular galaxy with
a merging protogalaxy  (see also K00). The effects  of bandshifting on
elliptical and S0 galaxies are  less pronounced than on spirals: their
FUV   emission  is   smoothly  distributed   but  is   more  centrally
concentrated  than  optical  light (\cite{oco99}).   Although  imaging
distant galaxies in the  infrared would lessen bandshifting effects by
sampling the rest--frame  optical instead of the UV  out to $z\sim 4$,
the  rate of  NIR data  acquisition is  currently slower  than  in the
optical.  Deep NIR  imaging of small areas using  the NICOMS camera on
HST suggests that  the increase in peculiarity may  not necessarily be
due to bandshifting (Teplitz et  al.  1998, \cite{bun99} and Corbin et
al. 2000),  which could be  confirmed more rapidly with  large optical
samples and a better understanding  of the relationship between UV and
optical  morphology.  Our  main point  is  that the  vast majority  of
publications  to  date  on  high-redshift morphology  are  subject  to
band-shifting effects, hence the present work is important in light of
re-interpreting those results.

\par  In the  absence  of data,  several  attempts have  been made  to
simulate    the    appearance    of   high--redshift    galaxies    by
``extrapolating'' the UV morphology from optical images or by applying
the  cosmological effects of  surface brightness  dimming and  loss of
spatial  resolution.   Artificial  UV  images of  galaxies  have  been
produced  using  template  spectral  energy distributions  (SEDs)  for
different Hubble types to estimate the UV light in each pixel based on
its optical  colors (\cite{abr96a}; \cite{afm97};  \cite{bri98}).  The
apparent  change in  morphology due  to bandshifting  in  the $A$--$C$
classification system has been  quantified by Brinchmann et al. (1998)
using this approach.  They find that 13\% of spirals are mislabeled as
irregular/peculiar at $z  \sim$ 0.7, with the fraction  rising to 24\%
at $z \sim$ 0.9.  This fraction  may be expected to increase at higher
redshifts as rest  wavelengths move further into the  UV.  However, it
is important to  note that these simulations do  not utilize actual UV
data.   The SEDs  used  to infer  UV  flux from  $B-R$  color may  not
accurately reflect local conditions,  especially in very dusty regions
or in  localized starbursts  (\cite{don95}).  Conselice et  al. (1999)
degrade the  resolution and signal--to--noise (S/N)  of optical images
of nearby galaxies to simulate other effects of distance and redshift.
They find their technique for  measuring asymmetry to be robust to S/N
variations  (for S/N~$>$~100)  but to  show  an apparent  drop in  the
asymmetry index as spatial resolution decreases.  Working with UV data
directly to  avoid uncertainties in  the morphological $k$--correction
due to bandshifting, \cite{gia96} and \cite{hib97} rebinned images and
scaled  the   surface  brightness  to   simulate  redshifting.   Their
qualitative analysis of the resulting morphology suggests that even in
the   absence  of   bandshifting,  cosmological   effects   cause  the
``redshifted''  galaxies  to have  a  later  type  and more  irregular
appearance (Giavalisco et al.  1996, Hibbard \& Vacca 1997).  However,
without objective criteria to  describe morphology, it is difficult to
determine  whether  the magnitude  of  this  effect  is sufficient  to
account  for  the observed  increase  in  irregular  galaxies at  high
redshift.  (We note in passing that there is a contradiction regarding
the  trend of  ``type'' based  on  symmetry with  decreasing S/N  when
comparing the results of Conselice  et al. with those of Giavalisco et
al. and Hibbard \& Vacca.  Reconciling these differences is beyond the
scope of this paper but the referee suggested that we alert readers to
this fact).  \cite{gar97} have  measured the central concentration and
asymmetry of five  nearby galaxies on UV images  resampled to simulate
$HST$ images at  redshifts near $z \sim$2, but  the effects of surface
brightness dimming  were not considered.  They also  find that spirals
tend  to move  to  a later  type  bin in  the $A$--$C$  classification
system.  A quantitative study of  the change in apparent morphology of
local  galaxies when  viewed at  high redshift  is still  necessary to
determine how the  distant galaxy population differs from  that of the
current epoch.

\par  In order  to  compare the  morphologies  of galaxies  at UV  and
optical wavelengths,  we observed a set  of 32 nearby  galaxies in the
FUV using the Ultraviolet  Imaging Telescope (UIT) in low--earth orbit
aboard  the Space  Shuttle  Endeavour.   In K00,  we  present FUV  and
ground--based  optical  data  for  these galaxies  and  discussed  the
morphology  qualitatively  based  on  images  and  surface  brightness
profiles.  We find that localized star formation features dominate the
FUV light and  FUV--optical color profiles, in contrast  to the smooth
underlying  distributions seen  in optical  images and  profiles.  The
strong deviations from a smooth disk  or disk + bulge profile that are
evident in the FUV highlight the difficulty of determining traditional
structural parameters from rest--frame UV images and suggest a need to
move beyond the Hubble sequence to describe UV galaxy morphology.

\par  In  this  paper,  we  present a  quantitative  analysis  of  the
morphology of the K00 galaxy sample.  We use central concentration and
asymmetry parameters to quantify bandshifting and cosmological effects
that are inherent in the  study of high--redshift galaxies.  This work
extends the  qualitative investigations of cosmological  effects on UV
images  carried  out by  \cite{gia96}  and  \cite{hib97}.  Unlike  the
Brinchmann et al.  (1998) study of redshift effects  on morphology, we
do not  rely on  the ``extrapolation'' of  UV morphology  from optical
images.  Section~2  reviews the sample selection  and data acquisition
and  reduction, most  of  which is  discussed  in detail  in K00.   In
Section~3,   we  present   our  method   for  measuring   the  central
concentration   and   asymmetry.   The  procedure   for   artificially
redshifting images  to take into  account dimming and  reduced spatial
resolution  is  presented  in  Section~4,  along  with  a  qualitative
discussion  of  these   effects  on  apparent  morphology.   Section~5
contains  the results of  the quantitative  morphology analysis  and a
discussion of the total  morphological $k$--correction for $C$ and $A$
between  local  optical data  and  high--redshift  galaxy images.   In
Section~6,  we  provide  a  brief summary,  compare  our  artificially
redshifted galaxies  to deep HST images, and  discuss implications for
the study of high--$z$ galaxies.  We also consider the shortcomings of
both the Hubble scheme and the $C$ and $A$ indices for UV galaxy data,
and we propose  some new parameters that may  more adequately describe
the observed morphology.

As  emphasized  by the  referee  there  are  certainly other  ways  of
creating  asymmetry and concentration  indices that  do not  depend on
isophotes  and may  indeed  be  more robust  (to  centering errors  in
particular).  In this paper we  are testing the methodology of Abraham
et al. (2000) since this is representative of the extant literature on
the  morphology  of  high-redshift   (HDF)  galaxies.   Our  data  are
publically  available for  alternative  tests, such  as the  iterative
minimization  scheme as  described in  Conselice, Bershady  \& Jangren
(2000) and in Bershady et al. (2000).

\section{The Data}

\subsection{Sample Selection}
\par The morphological analysis in this paper was carried out on
galaxies drawn from the sample of UV--optical data presented in K00,
with the addition of a few galaxies for which there are UIT images but
no optical data in our database.  The sample selection for the UIT
$Astro-2$ mission, our source of UV galaxy images, is discussed in
detail in K00.  Although the UIT sample was designed to cover a range
of morphologies from elliptical to irregular/peculiar, it does not
statistically represent the local distribution of Hubble types.
Because we wish to investigate in particular the morphology of the
irregular faint galaxies seen at high redshift, our sample is weighted
towards the spiral and irregular UIT galaxies.  Different subsets of
the sample are used for various parts of our study: only galaxies that
fit entirely on the optical data frames and do not suffer central
saturation in the optical filters are used to probe the behavior of
apparent morphology with wavelength, and only galaxies that are still
detectable in the FUV after artificial redshifting are used for the
study of redshift--dependent effects.  We also artificially redshifted
the $U$--band images of some galaxies into longer wavelength optical
filters to study the effects of moderate redshift; again only galaxies
that fit entirely on the optical frame were used.  Table~\ref{galxdat}
contains some basic data for each galaxy in the sample and denotes for
which parts of our investigation they were used.  More detailed
information about most of these galaxies and our observational data is
given in K00.

\subsection{Observations and Data Reduction}
\par FUV images were obtained using the UIT and underwent the standard
pipeline processing (\cite{ste97}).  $U$, $B$, $V$, $R$, and $I$
images were obtained over the past several years at Palomar
Observatory, Las Campanas Observatory, and Cerro Tololo
Inter--American Observatory.  The CCD images were bias--subtracted,
flatfielded, and combined (where necessary) using standard procedures
in IRAF (\cite{tod86}, VISTA (\cite{sto88}), and CCDPACK in
Starlink. Artifacts such as the ``UIT stripe'' and cosmic rays on the
optical images were removed, and background sky levels were determined
for each image.  We transformed the coordinates of the optical images
to the system of the FUV image for each galaxy, yielding a final scale
of 1.14\arcsec/pixel.  The optical images were then smoothed to the
UIT resolution of $\sim $3\arcsec.  Finally, we masked out all
foreground stars in the optical data and interpolated over those
regions of the image.  Data reduction procedures are described in more
detail in Kuchinski et al. (2000)

\section{Concentration and Asymmetry Indices}

\par We have chosen to use the central concentration and asymmetry
indices as quantitative morphology parameters because they are
conceptually simple and have been used in a significant body of recent
work, especially in the study of distant galaxies (see references in
the Introduction).  In general, galaxies with high degrees of central
concentration and symmetry will have an ordered, regular appearance
and those with low central concentration and large asymmetry will have
an irregular or peculiar morphology.  (Here we explicitly distinguish
between the Hubble ``Irregular'' type that is simply an extension of
the late--type spiral sequence and the generic ``Peculiar'' type that
has merger or interaction signatures, or unusual features that do not
correspond to any other Hubble types.)  The combination of these two
parameters, plotted as log~$A$~$vs.$~log~$C$, has been shown by
Abraham et al. (1996b) to divide nearby ( optically--classified)
galaxies into three broad morphological bins corresponding to the
Hubble sequence: elliptical/S0, spiral, and irregular/peculiar/merger.
This provides a basis for understanding distant galaxies in terms of
the well--studied (at optical wavelengths) local population.  As noted
in several instances below, the intrinsic patchiness and low
signal--to--noise of the FUV images have posed special difficulties
for the techniques typically used to measure $C$ and $A$.  We have
therefore modified the definitions proposed by Abraham et al. (1994,
1996), while retaining the general idea of a correlation between high
concentration plus symmetry and an ordered appearance.  The effects of
these modifications will be explored later in this section.

\par  It is  important to  carefully  define a  galaxy's position  and
aperture  in which  the concentration  and asymmetry  indices  will be
measured  because these  indices  are  known to  be  sensitive to  the
centering  and aperture  size (\cite{tep98};  Conselice et  al. 1999).
Many  authors define the  aperture based  on a  threshold set  at some
small   multiple   of  the   measured   sky   noise,   and  then   use
intensity--weighted  image moments  to define  an ellipse  ({\it e.g.}
\cite{abr96b};  \cite{tep98}). Alternatively,  we note  that  there is
also   the  method   of  Bershady   et  al.    (2000)  which   is  not
isophote-threshold dependent.  Aperture centers are typically obtained
by centroiding  on the  brightest pixel (\cite{tep98}).   However, the
patchy appearance  and prominence of star--forming regions  in the FUV
images poses some significant  difficulties for using these techniques
in  an  unmodified manner.   Sigma--clipping  often produces  multiple
small regions within what visually  appears to be the galaxy aperture.
Spiral arms or  large areas of star formation can  in fact be brighter
than the  geometric center  of the galaxy  (as defined by  the optical
center or by the apparent  center of an ellipse enclosing the galaxy).
Here  we have  used the  available optical  data to  our  advantage in
defining  the apertures.  (It  should  be noted  that  this would  not
usually be possible  in analyzing high--redshift galaxies).  Isophotal
fits to  outer parts of  the longest wavelength galaxy  image (usually
$R$ or  $I$; images of  the same galaxy  in both filters  yield nearly
identical results) determine the ellipticity and position angle of the
aperture.  The centroid  of this long wavelength image  is selected to
be the fixed aperture  center.  These ellipse parameters are identical
to  those  used in  K00  for  azimuthal  averaging to  obtain  surface
brightness  profiles.   The maximum  aperture  radius  is the  largest
radius at which there is still detectable light above the noise in the
FUV  image, which  was determined  by visual  inspection of  the light
profiles of K00.  This  procedure for aperture radius selection yields
a  slightly  different limiting  surface  brightness  for each  galaxy
(depending   on  S/N),  which   will  be   considered  below   in  our
determination of uncertainties  in $C$ and $A$.  As  a control, we use
the same aperture  for each filter's image of a  galaxy to ensure that
we compare the same physical region of that galaxy at all wavelengths.
For the small  number of galaxies for which we have  only FUV data, we
estimate  the  center  and   determine  the  ellipse  parameters  from
isophotal  fits  to  the  FUV  image.   In all  of  these  cases,  the
ellipticity agrees well with the $RC3$ value (\cite{dev91}), and these
particular  galaxies are  regular  enough that  the  center is  fairly
unambiguous.   The ellipse  parameters for  each galaxy  are  given in
Table~\ref{cadat}.  As in K00, the  merger NGC 4038/9 is considered as
a single system  with a center determined from  a segmented image, and
the aperture for  NGC 3227 also includes light  from its companion NGC
3226.

\par The concentration index $C$ is the fraction of total galaxy light
that is emitted from the central region compared to the whole.  In
practice, different authors have measured $C$ using a variety of
definitions (Doi et al. 1993; \cite{abr94}; \cite{nai97b}).  We select
a very simple expression:
\begin{eqnarray}
C & = & \frac{\Sigma_{(i,j:\; R<0.3R_{\rm max})} I(i,j)}{\Sigma_{(i,j:\; R<R_{\rm max})} I(i,j)}
\end{eqnarray}
where $R_{\rm max}$ denotes the radius of the elliptical galaxy
aperture selected as described above.  The calculation of total flux
in the inner and outer elliptical apertures is performed on
background--subtracted galaxy images.  The choice of what fraction of
$R_{\rm max} $ constitutes the central region is somewhat arbitrary;
for convenience of comparison, we select the value of 0.3 that was
adopted by Abraham et al. (1994) and used in most subsequent analyses
of galaxy concentration by these and other workers.  Noise in the
images should not present a serious problem for our concentration
index because it should be evenly divided between ``positive'' (above
the sky) and ``negative'' (below the sky) and thus should cancel out
within each aperture. Concentration indices for each galaxy at every
available wavelength are given in Table~\ref{cadat}.

\par The asymmetry index $A$ is a measure of the 180$^{\circ } $
rotational symmetry of the galaxy.  The basic mathematical definition
is as follows:
\begin{eqnarray}
A & = & \frac{\Sigma_{(i,j: \; R<R_{\rm max})} |I(i,j)-I_{\rm rot}(i,j)|}{\Sigma_{(i,j: \; R<R_{\rm max})} I(i,j)} 
\end{eqnarray}
where the original image is rotated 180$^{\circ } $ to get $I_{\rm
rot}(i,j)$.  By taking the absolute value of the difference between
the original and rotated images, one introduces a systematic error
because the sky noise always contributes positive values.  For
galaxies with regular shapes, in which all pixels in the aperture
contain galaxy light, this error is a small fraction of $A$.  On FUV
images, which have low signal--to--noise and are intrinsically patchy
or fragmented, the noise may be significant.  Abraham et al. (1996b)
correct for the noise by subtracting from $A$ the measured asymmetry
of a patch of sky with identical size to the aperture; but we
typically do not have enough sky on our galaxy frames to apply this
corrective technique.  Instead, we consider the sky noise in pixels
that do contain galaxy light, and we also explicitly take into account
the fact that many pixels within the aperture do not contain any
galaxy light. We approach the latter problem by summing over only
those pixels with values above a threshold of $1.5\sigma_{\rm sky} $
above the background, rather than all pixels in the aperture, when
using Equation~(2) for $A$. For optical images, most or all of the
pixels are used in the sum, while in the FUV, the fraction may be as
low as 10\%.  To correct for sky noise in the pixels that {\it do}
contain galaxy light (and thus are used in the sum), we measure the
asymmetry in a patch of sky as large as we can find on the image, then
renormalize it to the number of pixels actually used in the sum.  This
correction factor is then subtracted from the asymmetry determined for
pixels above the threshold.  Thus the noise--corrected expression for
the asymmetry is given by:
\begin{eqnarray}
A_{corr} & = & \frac{\Sigma_{(i,j:\; p(i,j)\;>\;n\sigma )} |I(i,j)-I_{\rm rot}(i,j)|}{\Sigma_{(i,j:\; p(i,j)\;>\;n\sigma )} I(i,j)} \;-\; k_{A}
\end{eqnarray}
where $p(i,j)$ is the pixel value at position $(i,j)$, $n\sigma $ is
the threshold, and $ k_{A} \; = \; N_{pix} * A_{sky, pix} $. $N_{pix}$
is the number of pixels in the aperture with values above the
threshold, and $A_{sky, pix}$ is the asymmetry per pixel due to sky
noise, calculated by dividing $A$ of Equation~(2) for a patch of sky
by the area (in pixels) of the patch.  Values of the correction term
$k_{A}$ range from $\leq 0.03$ for optical data to 0.1--0.2 for FUV
images ($A$ itself ranges from 0 to 1).  The galaxy asymmetry values
calculated with this method are given in Table~\ref{cadat}.

\par We estimate the uncertainty in $C$ and $A$ by considering two
factors: measurement error and errors that arise from using apertures
extending to different limiting surface brightnesses.  To quantify the
error in measurement, we use five galaxies common to our sample and
the digital atlas of Frei et al. (1996).  (A sixth galaxy, M63, was
saturated in the Frei et al. images and was not used for the
comparison.)  We calculate $C$ and $A$ from both sets of images (ours
and theirs) and find $ \sigma_{C} = 0.016$ and $ \sigma_{ A} = 0.011$.
The limiting surface brightness on the FUV images, which was used to
determine the aperture radius for measuring $C$ and $A$, ranges from
$\sim 24 - 25$ mag arcsec$^{-2}$.  For nine galaxies with limiting
surface brightness $\geq 24.5$ mag arcsec$^{-2}$, we measured $C$ and
$A$ in an aperture extending only to 24 mag arcsec$^{-2}$ and compared
the values to those measured in the maximum aperture size.  In this
case, we find $\sigma_{C} = 0.025$ and $\sigma_{A} = 0.012$.  The
direction of the changes in $C$ and $A$ as aperture radius increases
depends on the detailed structure of the galaxy, but it is typical to
find a larger concentration and a larger asymmetry in the larger
aperture.  Adding these (independent) errors in quadrature yields our
final uncertainty estimates: 0.03 for $C$ and 0.02 for $A$.

\par We have also used the  Frei sample to explore the effect of using
different mathematical  definitions of the $C$ and  $A$ indices.  This
sample  was used  by  Abraham et  al.  (1996) to  calibrate the  $A-C$
classification system,  and their $C$  and $A$ indices  for individual
galaxies are  tabulated.  We first  tested our ability to  recover the
Abraham et al.  values by using {\it their} definitions of $C$ and $A$
to measure indices on the Frei sample.  The results are encouraging: a
mean difference in  $C$ of $0.001 \pm 0.015 (1 \sigma)$  and in $A$ of
$0.006 \pm 0.047$, both in the sense of (our measurement -- Abraham et
al.).  We then  used {\it our} definitions of $C$ and  $A$ on the Frei
sample to quantify the  systematic difference between the two methods.
The measurement of $C$ differs only in the aperture definition, and we
find an offset of $\Delta  C({\rm ours-Abraham}) = 0.03 \pm 0.02$.  In
spite of the very different methods used to measure $A$, the offset is
still small:  $\Delta A({\rm ours-Abraham})  = 0.04 \pm  0.03$.  These
offsets are small compared to the uncertainties that we have estimated
above, and small  with respect to the errors of  $\sim 0.07$ quoted by
Abraham  et  al.\footnotemark[4]  \footnotetext[4]{We  note  that  the
scatter in  our calculated offsets  is somewhat small compared  to the
uncertainties: this is likely because  we have used the same images as
Abraham  et al.  did.  The  quoted uncertainties  include a  term that
takes  into account  differences  in limiting  surface brightness  for
different galaxies,  which will not  be a factor in  comparisons using
the same image of the same  galaxy.}  Our results suggest that it will
be possible to utilize the Abraham et al.  $A-C$ classification scheme
using our values of $C$  and $A$, with the morphological bins adjusted
by the calculated  offsets where appropriate.  For an  analysis of the
effects of different aperture sizes see Conselice, Bershady \& Jangren
(2000) who  have   explored  this   source  of  uncertainty   in  some
detail. Also see  Bershady, Lowenthal \& Koo (1998)  for a description
of measuring sizes for faint sources.

\section{Artificial Redshifting}
\par In order to study the influence on morphology of cosmological
effects associated with large look--back times, we have artificially
redshifted galaxy images to values of $z$ at which the rest--frame FUV
filter bandpass would coincide with the four $HST$ WFPC2 filters used
to image the HDF (\cite{wil96}).  We simulate the HDF because it
comprises the deepest observation of distant galaxies to date, and has
been the subject of numerous studies of morphology at high redshift.
The FUV rest wavelength ($\sim $1500$\AA$) is redshifted into the
broad--band F300W filter at $z\sim 1 $, the F450W filter at $z\sim 2
$, the F606W filter at $z\sim 3 $, and the F814W filter at $z\sim 4 $.
(The precise value of $z$ depends on which FUV filter was used for the
UIT imaging and may vary from the approximate redshift given by up to
$\pm $0.2; details of the two FUV filters are given in K00.) Because
there is at present a paucity of $U$--band images of nearby galaxies,
we also explore the effects of moderate redshifts that move the
$U$--band rest wavelength into the F606W filter at $z\sim 0.6$ and
into the F814W filter at $z\sim 1.2$.  We emphasize that there is no
need to estimate the effects of bandshifting in these simulations; we
simply selected a redshift such that the rest wavelength is placed
directly into the desired filter.  For example, a high--redshift
galaxy at $z\sim 3$ observed with WFPC2 in the F606W filter is really
being observed in its rest--frame FUV, and a moderate redshift galaxy
at $z\sim 0.6$ in F606W is really being observed in its rest--frame
$U$--band.  By comparing the artificially redshifted images to their
unredshifted counterparts, we can isolate the effects of surface
brightness dimming and loss of spatial resolution on apparent
morphology.  Although this method restricts our investigation to
specific redshifts, we span the range of $z \sim 0.6 - 4$ over which
recent work has suggested intrinsic evolution in the galaxy
morphologies.

\par We follow the procedures outlined by \cite{gia96} to artificially
redshift the galaxy images, using their Equations~(2),~(5),~and~(7) to
determine  the resampling  factor and  surface brightness  scaling.  A
cosmology with  $\Omega=1$ and $q_{0}=0.5$ is  assumed throughout, and
we   take  $H_{0}\;=\;75   $  km   s$^{-1}$  Mpc$^{-1}$   ({\it  e.g.}
\cite{fre01}).   Because  this  value  of  $H_{0}$ was  also  used  to
estimate  the distances  to sample  galaxies, both  the  rebinning and
brightness  scaling  factors  will  be independent  of  $H_{0}$.   The
dependence on $q_{0}$ is much more complex, but Giavalisco et al. test
values of  both 0.05 and  0.5 and find  that galaxies are  more easily
detected with  the higher $q_{0}$.  Interested  readers should consult
this paper for details.   Filter and detector characteristics that are
input to these  formulae are taken from Stecher et  al. (1997) for the
UIT and Williams et al.  (1996), {\it The WFPC--2 Instrument Handbook}
(Biretta 1996),  and {\it The HST  Data Handbook} (Voit  1997) for the
HDF.  Galaxy distances are from  \cite{tul88} and are given in Table~1
of K00.   Instead of  convolving the resulting  images with  the $HST$
point spread function (PSF) as  Giavalisco et al. have done, we choose
the more simple method of  smoothing with a circular Gaussian to match
the PSF  width of  $\sim $3  pixels reported by  Williams et  al. This
technique avoids  the difficulty of  simulating in detail  the complex
PSF that results from $HST$  optics and the ``drizzle'' procedure used
to combine HDF images.  Our  redshifted images were then scaled to the
appropriate HDF  exposure time for  each filter, and a  sky background
and sky noise were added based  on the HDF values reported in Williams
et al.  We accounted for foreground Galactic extinction (very small in
most cases)  by correcting the  input image to zero  extinction before
redshifting, then  adding the  appropriate extinction for  each filter
after  the  artificial   redshifting  was  completed.   The  $B$--band
foreground extinctions for each  galaxy are from \cite{bur84}, and the
Galactic  extinction  law  of   \cite{ccm89}  was  used  to  calculate
extinctions at other wavelengths.

\par As galaxies are artificially redshifted to high $z$, their
appearance can change dramatically simply due to the fading of low
surface brightness features and simultaneous loss of spatial
resolution.  In some systems, only the bright regions may survive the
effects of dimming.  Observed at simulated $z \sim 4$ in the F814W
filter of the HDF, the Im galaxy NGC 4214 appears in
Figure~\ref{fuvred} as a compact, regular object.  For other galaxies,
such as the Scd M101 (also shown in Figure~\ref{fuvred}), only bright
star--forming regions in the spiral arms are visible at high redshift;
the nucleus is quickly lost below the detection threshold.  The
spatial distribution of these features gives the artificially
redshifted images a patchy, fragmented appearance suggestive of a
later type galaxy, or even multiple systems (see also Giavalisco et
al. 1996a).  Most of the E/S0 systems in our sample are either
undetectable at high redshift or have shrunk to the appearance of
point sources. This strongly suggests that source counts at high
redshift will be affected by a lack of early--type galaxies unless
these systems have evolved significantly.  At lower redshifts ($z \leq
1$), cosmological effects are mild and the simulated galaxies simply
look like fainter, more smoothed versions of their local counterparts.
Figure~\ref{ured} shows the results for the Sbc galaxy M51, whose
rest--frame $U$--band data has been redshifted into the WFPC2 F606W
and F814W filters.  The simulated F606W image looks a great deal like
the original $U$--band image, but by $z \sim 1.2$ in the F814W filter,
the galaxy is beginning to look less regular as the faint inter--arm
light falls below the detection threshold.  It is clear from these
images that surface brightness effects play a significant role in
determining the apparent galaxy morphology, and thus that cosmological
dimming cannot be neglected in an analysis of high--redshift objects.

\par In the next section we will attempt to quantify the effects of
redshift on morphology using the concentration and asymmetry indices.
We consider only those galaxies that are detected and resolved on the
simulated high--redshift images.  First, $C$ and $A$ were measured on
the artificially redshifted images using the procedures described in
Section 3 and a fixed ``physical'' aperture size ({\it i.e.} size in
kpc on the galaxy).  For these measurements we select an aperture size
on the artificially redshifted image, then scale it by the appropriate
bin factor to determine its size on the unredshifted image.  This
procedure isolates the effects of surface brightness dimming and
reduced spatial resolution by comparing the same physical region of
the galaxy at different simulated redshifts.  We then recalculated $C$
and $A$ using a method that mimics the procedure an observer might
follow: adjusting the aperture size for each image by visual
inspection to enclose only the detectable light.  In this case, the
physical aperture size varies based on the limiting surface brightness
of each simulated image.  The apertures selected for high redshifts
are often smaller than the size expected from scaling a galaxy's
rest--frame aperture by the bin factor for that redshift.  This
comparison is less direct than the one in which aperture size is
fixed, but it is more indicative of analysis techniques for deep
surveys.  Table~\ref{redca} gives the $C$ and $A$ values for the
artificially redshifted FUV and $U$--band images in which apertures
were adjusted based on limiting surface brightness.  Differences
between these results and the values for a fixed aperture size are
discussed in Section 5.3.

\section{Quantitative Morphology Results}

\par In this section we discuss the results of our quantitative
morphology investigation.  We first consider the effects of
bandshifting alone by comparing the FUV and optical morphologies of
the sample galaxies within a fixed aperture (Section 5.1).  The
effects of surface brightness dimming and loss of resolution at high
redshift, which we shall refer to collectively as the ``cosmological
effects'', are then discussed in terms of a fixed physical aperture
size for each galaxy, {\it i.e.} one that samples the same physical
region on rest--frame and simulated--redshift images (Section 5.2). We
then compare the fixed aperture results with those obtained by
adjusting the aperture based on each image's limiting surface
brightness, as an observer would do with real data (Section 5.3).
Finally, we combine the bandshifting and cosmological effects to
discuss the total ``morphological $k$--correction'' between the
optical appearance of nearby galaxies and images of high--redshift
systems (Section 5.4).  In several of the plots of concentration and
asymmetry indices that are presented in this section, galaxies are
divided into four bins by Hubble type.  Symbols for the different bins
are explained in the figure captions.  Note that not all galaxies have
optical data in every filter; that is, a plot comparing the $B$-band
and FUV morphologies will not necessarily include all of the same
galaxies as one comparing $R$-band and FUV images.

\subsection{Bandshifting}
\par FUV values of the central concentration index ($C$) are
predominantly lower than those measured on optical images of the same
galaxy.  The left--hand panel of Figure~\ref{cawave} illustrates the
behavior of $C$ as a function of wavelength for ten representative
galaxies in our sample; galaxy names are labeled to facilitate the
discussion below.  The change $\Delta C\;=\;C_{\rm OPT} - C_{\rm FUV}$
is shown as a function of $C_{\rm FUV}$ for all galaxies in
Figure~\ref{delca}.  Galaxies with large increases in $C$ from FUV to
optical wavelengths, ({\it i.e.} large positive values of $\Delta C$),
such as NGC 1097 and NGC 1566, tend to be early to intermediate--type
spirals whose prominent optical bulges are faint to invisible in the
FUV (see also K00).  This effect is enhanced in barred galaxies due to
the dominance of red stars in the bar, as shown for the SBa galaxy NGC
1512 in Figure~\ref{bandsh}.  The change in $C$ is less dramatic for
late--type spirals such as NGC 1313, M51, and M63.  Galaxies with
ongoing, widespread star formation often have similar $C_{\rm OPT}$
and $C_{\rm FUV}$ because the light in both spectral regimes is
dominated by the young stars.  These systems include the starburst NGC
3310 (\cite{smi96}, also shown in Figure~\ref{bandsh}), and the
irregulars NGC 4214\footnotemark[5] \footnotetext[5]{\cite{fan97b}
find a starbursting core superposed on a faint disk in NGC 4214;
however, our aperture for $C_{\rm OPT}$ encloses only the region
containing FUV light and thus does not sample the disk.}  and NGC
4449, as well as the starburst~+~AGN NGC 1068 (\cite{nef94}).  These
galaxies have a higher $C_{\rm FUV}$ than others in our sample (see
Figure~\ref{cawave} for NGC 3310 and NGC 4214), suggesting that the
concentration index may be useful as an indicator of starburst
activity in the analysis of rest--frame FUV galaxy images.
Conversely, the merger system NGC 4038/9 shares the starburst trait of
similar $C_{\rm FUV}$ and $C_{\rm OPT}$, but has uniformly low values
of $C$ corresponding to its peculiar morphology (Figure~\ref{cawave}).
The starburst galaxy NGC 5253 is a conspicuous counter--example to the
trend toward lower $C$ in the FUV.  Its UV--bright, centrally
concentrated starburst produces a very high $C_{\rm FUV}$, while older
stars surrounding the burst contribute significantly to the optical
light and thus reduce the value of $C_{\rm OPT} $.  Overall, the
change in $C$ due to bandshifting ranges from near zero to a maximum
of $\sim 0.4$, or up to 40\% as $C$ theoretically ranges from 0 to 1.

\par The asymmetry values $A_{\rm FUV}$ are consistently higher than
$A_{\rm OPT}$, with a marked trend towards larger $\Delta A \; = \;
A_{\rm OPT} - A_{\rm FUV}$ in galaxies that are very asymmetric in the
FUV (see the right--hand panels of Figures~\ref{cawave} and
\ref{delca}). UV--bright circumnuclear star formation in a ``broken
ring'' shape dominates the FUV light of several galaxies, including
NGC 1097 (Figure~\ref{cawave}), NGC 1512 (Figure~\ref{bandsh}), and
NGC 3351.  This produces large $A_{\rm FUV}$ values even though the
galaxies appear symmetric in optical light.  In others, such as NGC
925 and NGC 2403, the underlying diffuse disk light is invisible in
the FUV, and patchy light from young stars dominates $A_{\rm FUV}$.
The merger system NGC 4038/9 stands out in Figure~\ref{cawave} due to
its extremely high $A$ at all wavelengths compared to the other sample
galaxies.  As was the case for the concentration index, the asymmetry
indices of the global starburst NGC 3310 (see Figure~\ref{cawave}) and
the starburst~+~AGN NGC 1068 are not very dependent upon wavelength.
With the exception of NGC 4038/9 (which is plotted to the far right as
an asterisk but is excluded from the fitting procedure described
below), the right--hand panels of Figure~\ref{delca} show a tight
correlation between $\Delta A$ and $A_{\rm FUV}$.  The diagonal lines
in Figure~\ref{delca} show linear least--square fits to the data;
solid lines are the fits to data for each filter individually, and the
dotted line is the fit to all data together.  The two lines in each
panel are nearly identical, so $A_{\rm OPT}$ can be predicted from
$A_{\rm FUV}$ using the dotted--line relation:
\begin{eqnarray}
\Delta A \: = \: A_{\rm OPT} - A_{\rm FUV} & = & -0.92(\pm 0.05) \: \times \: A_{\rm FUV} \: + \: 0.12(\pm 0.02)
\end{eqnarray}
where the standard deviation of the residuals is $\pm $0.07.  The
maximum value for $\Delta A$ can be quite large, reaching $\sim 0.7$
(70\%)in some cases.  We stress that this relation, while useful over
the range of Hubble types in our sample of local galaxies, needs
further testing of its validity, especially for the mergers and
protogalaxies encountered at high redshift.

\par We find one particular galaxy, NGC 4736, that highlights the
limitations of using $C$ and $A$ indices for morphological
comparisons.  FUV and $V$--band images of this galaxy are shown in the
bottom panels of Figure~\ref{bandsh}.  As can be seen in
Figure~\ref{cawave}, the values of $C$ and $A$ for NGC 4736 change
very little with wavelength in spite of dramatic morphological
differences between the FUV and optical images.  A bright
star--forming ring produces high concentration in the FUV, while a
large bulge increases the optical value of $C$.  Symmetry in the ring
that dominates the FUV images yields a low value of $A_{\rm FUV}$,
while an equally regular disk and bulge shape are responsible for a
low $A_{\rm OPT}$.  The prevalence of star--forming inner and
circumnuclear rings in the FUV (K00 and references therein) suggests
that cases such as NGC 4736 may not be uncommon, and thus that the use
of $C$ and $A$ to compare rest--frame UV and optical images may not
adequately describe the detailed effects of bandshifting.

\par We now come to the most interesting general point of this study.
In marked constrast to the results for optical data, the FUV values of
$C$ and $A$ fail to segregate galaxies in our sample into the broad
morphological bins of the $A-C$ classification system of Abraham et
al. (1996b).  Figure~\ref{logca} shows both FUV and optical $C$ and
$A$ for the sample galaxies on the log~$A$~$vs.$~log~$C$ diagram.  The
dashed lines in Figure~\ref{logca} are the Abraham et al. (1996b)
divisions into morphological bins, (shifted by the offsets between our
values and theirs that were calculated in Section 3).  Errors in
log~$C$ and log~$A$ depend on the values of these indices (we simply
propagate the uncertainties in $C$ and $A$) and are shown on
representative locations of the two diagrams.  For the most part, the
optical data for our sample galaxies lie close to their expected
locations on this diagram.  Three notable exceptions are the irregular
starburst galaxies lying near the border between E/S0 and spiral
classes: NGC 4214, NGC 4449, and NGC 5253.  Thus it seems that the
Irr/Pec/Merger bin of Abraham et al. describes peculiar/merging
galaxies in our sample ({\it i.e.}  NGC 4038/9 and NGC 3226/7), but
not the irregulars with centrally concentrated starbursts.  In the
case of the FUV data, it is clear from the lower panel of
Figure~\ref{logca} that a majority of galaxies have landed in the
extreme Irr/Pec/Merger region and some have undergone
counter--intuitive shifts in location compared to their optical
classifications.  The peculiar galaxy NGC 5253 shifts to the earliest
type bin in the FUV because of its centrally concentrated and highly
symmetric starburst. The irregular galaxies NGC 4214 and NGC 4449
remain on the boundary between E/S0 and spiral in both optical and
FUV, contrary to the behavior of the latest type spirals that have
migrated from the Spiral to the Irr/Pec/Merger bin in the FUV.
Several early--type spirals are located in the upper left corner, with
very low $C$ and high $A$, where the asymmetry in each case is due to
a feature that is enhanced in the FUV but is considered a small or
secondary part of the optical morphology.  The three most extreme
examples are NGC 1097 and NGC 1512, whose asymmetric circumnuclear
star formation regions dominate the FUV light, and NGC 2841, whose
high inclination and dusty disk produce a ``lopsided'' appearance that
is likely due to extinction.  Our sample is not large enough to
provide more than individual examples of the problems associated with
classifying UV galaxies using $C$ and $A$. However, the FUV data in
Figure~\ref{logca} do not appear to suggest any natural way to divide
the log~$A$~$vs.$~log~$C$ diagram that unambiguously maps back to the
canonical (optical) Hubble type.

\subsection{Cosmological Distance Effects}
\par Our redshift simulations suggest that the concentration index is
fairly robust to distance--related effects out to $z \sim 3$, but it
appears to have a slight tendency toward decreased $C$ on high
redshift images in the $z \sim 4$ simulation.  The left--hand panels
of Figure~\ref{delzca} show histograms of $\Delta C = (C_{\rm rest} -
C_{\rm redshifted} $), giving the difference between our original
``rest--frame'' images (in either the $U$ or FUV band) and their
artificially redshifted counterparts.  We have selected a bin width of
0.10 for $\Delta C$ and $\Delta A$, commensurate with the
uncertainties estimated in Section~3.  As noted at the beginning of
this section, $C$ is calculated in a fixed physical aperture that
samples the same regions of the unredshifted and simulated high-$z$
images.  The median value of $\Delta C$ is $ \leq 0.01$ for $z \leq
3$, but rises to $\sim 0.06$ for the $z \sim 4 $ simulation.  The
potentially significant role of surface brightness effects is
demonstrated in the galaxies M101 and NGC 4258, which are the outliers
with $\Delta C \sim 0.2 - 0.3$ in the redshift simulations of
rest--frame FUV data (note that these galaxies are not detected in the
$z \sim 4$ simulation).  Star--forming regions in the outer parts of
the spiral arms are brighter than the inner galaxy.  While light from
the central regions is still visible in the unredshifted image and
thus contributes to the inner aperture of $C_{\rm rest}$, it has
virtually disappeared in the artificially redshifted images and thus
is not evident in the measurement of $C_{\rm redshifted}$ (see
Figure~\ref{fuvred} for M101).  Special instances such as NGC 4258 and
M101 can exhibit changes in $C$ of up to 30\% due to cosmological
effects, but overall, most galaxies in our restricted sample have
$\Delta C \leq 5\%$ for $z \leq 3$.

\par Cosmological  distance effects appear to  introduce added scatter
in the  measurements of $A$ with  no clear systematic  drift to either
larger  or  smaller  values  at   high  redshift.   As  shown  in  the
right--hand panels  of Figure~\ref{delzca}, the scatter  is larger for
the intrinsically asymmetric FUV images than for the $U$--band images,
even when artificially redshifted to  similar $z$.  The scatter in $A$
due to distance  effects ranges from 4\% (=$1  \sigma$ about the mean)
at moderate redshifts of $z \sim 0.6$ to 12\% at a redshift of $z \sim
4$.  A galaxy  will exhibit $A_{\rm rest} >  A_{\rm redshifted}$ ({\it
i.e.}  $\Delta  A >  0$) if faint  irregular features have  faded away
with distance and/or if the decrease in spatial resolution has blurred
asymmetric  features in the  rest--frame image.   Conversely, galaxies
may appear  to have a higher  asymmetry at high redshift  ($\Delta A <
0$)   if  the  brightest   features  are   asymmetrically  distributed
star--forming regions or spiral arms  that do not get blurred together
with  distance.  Thus  changes in  the  value of  $A$ at  cosmological
distances depend  on the  rest--frame morphology of  each galaxy  at a
level of detail more complex  than is usually considered in the simple
Hubble type.  For optical and UV images taken with currently available
instruments  this suggests  that the  sensitivity of  $A$  to distance
effects will limit its usefulness  in comparisons of the morphology of
local and distant  galaxy populations; however with better  S/N and by
observing in the near infrared these limitations may be ameliorated.

\subsection{Aperture Effects}
\par The  difference between the values  of $C$ and  $A$ measured with
fixed  apertures and  those obtained  by adjusting  the  aperture size
based  on  limiting  surface  brightness  are  typically  quite  small
($<$5\%), as  shown in  the histograms of  $\Delta (C,A)  = (C,A)_{\rm
fixed} - (C,A)_{\rm adjusted}$ in Figure~\ref{apcor}.  The bin size is
smaller  than  that  used  in   Section  5.2  because  we  expect  the
uncertainties to  be less: we  are comparing measurements made  on the
{\it  same} image  with different  limiting radii.   The concentration
index  is  more sensitive  than  the  asymmetry  to small  changes  in
aperture  radius, which is  not unexpected  because $A$  is calculated
from sigma--clipped  images in  which the faint  outer regions  of the
galaxy are  not sampled in either measurement.   The left--hand panels
of Figure~\ref{apcor}  show some tendency for $C_{\rm  fixed} < C_{\rm
adjusted}$, especially in  low--redshift simulations.  Fixed apertures
were  selected  on the  fainter  high--$z$  simulations  and thus  are
smaller than adjusted ones for  these images.  This means that less of
the bright galaxy ``core'' falls  within the inner aperture of the $C$
measurement (see Equation 1 in Section 3).  An extreme example of this
effect produces  $\Delta C \sim -0.3$  for the FUV image  of NGC 4736:
the bright starburst ring falls outside 0.3$R_{\rm max, \, fixed}$ but
inside  0.3$R_{\rm max, \,  adjusted}$.  Bright  star--forming regions
near $0.3 R_{\rm  max}$ in the FUV images of NGC  4258 ($\Delta C \sim
-0.2$) and M101 ($\Delta C  \sim -0.1$) produce similar effects, which
are enhanced for  these galaxies by the large  change in $R_{\rm max}$
between the  fixed and  adjusted aperture calculations.   Such effects
are  less  noticeable  in  the artificially  redshifted  data  because
discrete  bright  features are  blurred  by  the  decrease in  spatial
resolution.  Values  of $\Delta  A$ cluster around  zero, with  only a
handful of galaxies exhibiting  $A_{\rm fixed} < A_{\rm adjusted}$ due
to exclusion  of asymmetric  outer regions like  spiral arms  from the
smaller fixed aperture.  We note that there are no aperture effects in
the  $z \sim  4$ simulations  because they  have the  highest limiting
surface brightness: the fixed  apertures were selected on these images
and thus are identical to  the adjusted apertures.  Thus our asymmetry
index  $A$ is  quite robust  to aperture  effects for  the  changes in
$R_{\rm  max}$ that  occur as  galaxies  fade with  redshift, but  the
concentration index  $C$ may change  either slightly or  radically for
individual  galaxies   depending  on   the  details  of   their  light
distributions.

\subsection{Morphological $k$--corrections}
\par We  can combine the  results of the previous  investigations that
isolated   bandshifting,  cosmological,   and   aperture  effects   to
investigate  the  overall difference  in  observed optical  morphology
between a sample of nearby galaxies and one of high--redshift objects.
We emphasize  that we have  not simulated the possibility  of enhanced
star--formation in  galaxies at  large look--back times,  although our
sample  does  contain  some  local starbursting  systems.   We  simply
describe  the difference  between  a population  of galaxies  observed
locally and  that {\it same},  unevolved, population observed  at high
redshift,  both  in  the  optical  part of  the  spectrum.   For  this
comparison  we have  considered only  galaxies that  have $C$  and $A$
measured in  the optical  and either the  $U$--band or  FUV artificial
redshifting simulations.  This comprises  a total of 20 galaxies, five
of which have no $U$--band data  and two of which have no artificially
redshifted  FUV data  (due  to  lack of  detection  after the  surface
brightness   dimming).    Qualitative    examples   of   the   overall
morphological $k$--correction are given in Figure~\ref{3redgal}, which
shows galaxies as they would  be observed locally in the $V$--band and
as  they  would appear  at  moderate and  high  redshifts  in the  HDF
filters.   \par As  shown in  Sections~5.1--5.3,  bandshifting effects
dominate the  ``morphological $k$--corrections'' for  our quantitative
indices, changing $C$ by up to 40\%  and $A$ by up to 70\% compared to
typical  changes  $\leq  10$\%  for  distance  and  aperture  effects.
Although  $\Delta A$ due  to bandshifting  is correlated  with $A_{\rm
FUV}$, scatter  introduced by distance effects hinders  the ability to
predict the value  $A$ that would be measured  for a particular galaxy
viewed  at high  redshift.  The  $k$--correction  for $C$  and $A$  is
presented  via  the  log~$A$~$vs.$~log~$C$ classification  diagram  in
Figure~\ref{finlogca}.  Optical measurements for each galaxy are shown
in  the  top  left  panel,  followed  by  rest--frame  $U$--band  data
redshifted to $z \sim 0.6$ and $z \sim 1.2$, then rest--frame FUV data
redshifted to  $z \sim 2 - 4$.   Again, not all galaxies  are shown on
every panel due to differences  in the availability of data.  The data
in  Figure~\ref{finlogca}  show  a  clear trend:  at  each  succeeding
redshift, the bulk of the points move toward lower $C$ and higher $A$.
By   $z  \sim   2$,  most   of  our   sample  galaxies   lie   in  the
Irregular/Peculiar/Merger  region.  Galaxies that  do not  follow this
trend  are those  (already  noted in  Section~5.1)  that have  similar
optical  and FUV  morphology due  to their  prominent  starbursts: NGC
3310, and to a lesser degree  NGC 4214 and NGC 4449.  NGC 1068 retains
a very  high central concentration even at  high redshift (rest--frame
FUV)  due  to  the  starburst~+~AGN  core.   The  ``trajectories''  of
position  changes with redshift  in the  log~$A$~$vs.$~log~$C$ diagram
are shown  for several galaxies in  Figure~\ref{traj}, emphasizing the
difference between  spirals with dramatic changes  in morphology ({\it
e.g.}  NGC 1097)  and starburst  systems ({\it  e.g.} NGC  4449) whose
appearance would be similar locally  and at high redshift.  Given this
non--monotonic migration of data points with redshift, it is no longer
possible to divide this diagram into bins by Hubble type.  Many of the
early--type (Sa-Sb)  spirals have joined the ``mergers''  in the upper
left--hand  corner,  while later--type  spirals  and  Im galaxies  are
spread over  the ``early type'' regions  of higher $C$  and lower $A$.
These results  suggest that new  methods will be required  to classify
high--redshift galaxies:  UV values of  $C$ and $A$ are  inadequate to
infer the optical morphology because the spatial distribution of young
massive  stars  that  dominate the  UV  emission  may  or may  not  be
correlated with  that of older  stars that dominate the  optical light
and its morphological appearance.

\section{Summary and Discussion}

We  have  used numerical  parameters  that  individually describe  the
central concentration  and the  asymmetry of galaxies  to intercompare
their apparent  morphology in the UV  and optical and  to quantify the
effects  of redshift  on the  appearance of  distant  stellar systems.
Bandshifting and  distance effects, including changes  in the apparent
galaxy  aperture  as the  limiting  surface  brightness  of the  image
varies, are  each considered separately,  then combined in  an overall
discussion of the ``morphological $k$--correction'' in the $C$ and $A$
indices.  Unlike previous  models of the behavior of  $C$ and $A$ with
redshift ({\it  e.g.} \cite{abr96b}, \cite{bri98}), we do  not need to
extrapolate  the UV  galaxy image  from  optical data  because we  use
direct FUV images  from the UIT.  The inherent  patchiness of galaxies
in the UV  and the low signal--to--noise of our  data required that we
modify the traditional methods of  determining $C$ and $A$, which were
developed  for high  signal--to--noise  optical data  (\cite{abr96b}).
However, we note  that {\it a}) results from the  two methods are only
slightly offset,  with small scatter,  and {\it b}) the  comparison of
$C$  and  $A$ at  different  wavelengths  and  simulated redshifts  is
internally consistent  as it  was done using  the same method  for all
measurements.

\par  We find  that  bandshifting effects  dominate the  morphological
$k$--correction, while distance effects introduce small, mostly random
changes  in the  concentration or  asymmetry indices.   While  the FUV
values  of $C$  and $A$  generally show  an extension  of  the optical
trends (noted  by \cite{bri98}) toward lower  concentration and higher
asymmetry at short wavelengths, they  can be much larger in magnitude:
up to 40\%  for $C$ and 70\% for  $A$. Some counter--intuitive results
are produced  by the  strong dependence of  FUV morphology  on current
star--formation:  circumnuclear  rings  in some  early--type  spirals,
combined with the fading of  the red bulge at UV wavelengths, produces
the  high  $A$   and  low  $C$  which  are   usually  associated  with
irregular/peculiar  objects  in   spite  of  these  galaxies'  regular
appearance in the optical.   Centrally concentrated starbursts in some
late--type  and  irregular  galaxies  can  give them  a  more  regular
morphology in the  FUV, with the high $C$ and low  $A$ typical of E/S0
galaxies.  \par  Our simulations of  distance effects alone  show that
they tend  to produce small  changes (up to  7\%) in $C$  and slightly
larger offsets (up to 12\%)  in $A$.  These values are measured within
a fixed physical aperture (in kpc) that samples the same region of the
galaxy at  each redshift.  When the  aperture size is  varied to match
the  limiting  surface  brightness  of individual  images,  additional
scatter of  $\sim 3\%$ is  found in our  measurements of $C$  and $A$.
However, there are individual galaxies for which distance and aperture
effects are  particularly noticeable due to the  location of extremely
bright star--forming features well  away from the galaxy center. These
examples  provide  a  cautionary  note  on the  possibility  of  large
excursions in the derived type due to distance effects.

\par In this  paper, as well as  in K00, we find that  it is extremely
difficult  to  unambiguously map  a  galaxy's  FUV  morphology to  its
optical Hubble type, or to the $C$ and $A$ indices in the optical that
appear to be  correlated with Hubble type.  The  images in K00 suggest
that most  galaxies exhibit a  later Hubble type  in the FUV,  but the
magnitude of  this shift, even  in qualitative terms, is  clearly much
larger for some galaxies than for others.  The same phenomenon is seen
in  the   $A-C$  classification  system,   for  which  a   variety  of
morphological  $k$--corrections are  shown  in Figure~\ref{traj}.   We
show in Figure~\ref{finlogca} that the  overall changes in $C$ and $A$
with  redshift make  it nearly  impossible to  use  the classification
system of Abraham et al. (1996)  for rest--frame FUV data, or beyond $
z \sim  2$ in our simulations.   Moreover, it is difficult  to see how
the three  morphological bins  based on optical  Hubble type  could be
redrawn to accommodate  the rest--frame FUV data.  The  location of UV
galaxies  on the log~$C$--log~$A$  diagram depends  on details  in the
spatial pattern  of star--formation ({\it  e.g.} circumnuclear rings),
which are not necessarily included in the simple Hubble type.

\par One  of the nagging  questions in the interpretation  of peculiar
morphologies in the HDF and other  deep surveys is whether or not they
are typical spiral or irregular galaxies viewed at high redshift ({\it
e.g.}  \cite{bun99}).   The best  available  comparison  data for  our
simulated  sample  of  galaxies  at  cosmological  distances  are  the
$A$--$C$  classifications  of HDF  galaxies  presented  by Abraham  et
al.   (1996a).    Figure~\ref{hdfplot}   shows   the   loci   on   the
log~$A$~$vs.$~log~$C$ diagram for our redshifted simulations (points),
the  HDF morphologies  from Abraham  et al.  (shaded polygon)  and the
local optical galaxy sample  of \cite{fre96} artifically redshifted to
$z = 0.5 - 2.0$ by Abraham et al. (open polygon).  The shaded and open
polygons have  been shifted from their  loci in Abraham et  al. by the
offsets  calculated in Section~3.   Nearly all  of our  simulated data
points for  $z \leq  1.2$ lie  within the open  polygon, which  is the
locus  of   non--evolving  galaxies.   The  lack  of   points  in  the
right--hand half of  the polygon is due to  the paucity of ellipticals
in our sample.  It is likely that  our $z \sim 2$ data lie outside the
open polygon  because the  rest--frame FUV of  our simulations  is not
analogous  to the  rest  wavelength of  the  Abraham et  al.  $z =  2$
simulation.  Our  artificial $z  \sim 2$ data  are rest--frame  $ 1500
{\rm \AA} $  images viewed through the F450W  filter, while Abraham et
al. (1996a) use  the F814W HDF images, which  sample rest--frame $\sim
2500 {\rm \AA} $ at $z \sim 2$.   (We have no data at $ 2500 {\rm \AA}
$ due  to the failure of  the mid--UV camera during  the UIT $Astro-2$
mission.)  If we  assume that galaxy morphology at $  2500 {\rm \AA} $
is  intermediate between  that  at  $U$--band and  FUV  (see also  the
mid--UV images in Marcum et al.  1997, 2001), we can expect that their
$A$--$C$ locus  would be  in the  upper left part  of the  HDF region,
between  our $z  \sim 1.2$  and $z  \sim 3$  simulations.   Abraham et
al.  interpret  the  HDF  galaxies  that  lie  outside  the  locus  of
artificially redshifted  Frei data as  either very late  Hubble types,
spirals   that   have  evolved   since   this   look--back  time,   or
mergers/peculiar galaxies.   However, the  presence of several  of our
simulated $z \sim  3$ points within that part of  the shaded HDF locus
may suggest  that these  HDF galaxies  have $z \geq  3$ and  are being
sampled in their rest--frame UV  even on the F814W images.  Our sample
size is  insufficient to speculate on what  fraction of high--redshift
peculiar galaxies might  be similar to local spirals,  but our FUV $C$
and  $A$ measurements  demonstrate that  it is  certainly  possible to
reproduce  examples  of  the  apparent peculiar  HDF  objects  without
necessarily invoking dramatic evolution.

\par  Our  artificially  redshifted  FUV galaxy  images  also  provide
qualitative  support  for  the  idea that  bandshifting  and  distance
effects alone  are capable of  producing the morphologies  observed in
high--redshift galaxies.  The  top left panel of Figure~\ref{peculiar}
shows the Sab galaxy  NCG 1068 as it would be observed  at $z \sim 4$.
Its  FUV--bright starburst  + AGN  center  gives the  appearance of  a
compact  core  surrounded  by   diffuse  light.   This  morphology  is
strikingly  similar   to  that  observed  by   \cite{gia96b}  in  many
Lyman--break galaxies, and the approximate radius of the NGC 1068 core
at  this redshift  ($\sim 0.4$  arcsec)  is comparable  to their  core
radii.  The  highly inclined Sd galaxy  NGC 4631, when  observed at $z
\sim 3  $, (see Figure~\ref{peculiar}), bears a  strong resemblance to
the  ``chain galaxies''  presented in  Figure~20 of  \cite{cow95}.  We
estimate  an FUV  axial ratio  of $\sim  5$ for  this galaxy  based on
ellipse fitting to the rest--frame  image, which is near the middle of
the  range quoted  by Cowie  et al.   At increasing  redshift, surface
brightness effects cause the galaxy to appear even thinner and produce
the appearance of discrete star--forming regions ``chained'' together.
The   FUV  images  of   NGC  4631   support  the   interpretations  of
\cite{dal96},   \cite{van96},  and   \cite{smi97}  that   the  ``chain
galaxies'' are  in fact edge--on  disks.  Other peculiar  galaxy types
noted  in  the HDF  by  van  den Bergh  et  al.   are the  ``tadpole''
(head--tail) galaxies  and several  examples of multiple  mergers (see
also \cite{pas96} for a discussion of the mergers of compact objects).
At   $z  \sim   3$,   the  Sd   galaxy   NGC  1313   (also  shown   in
Figure~\ref{peculiar}) could easily be mistaken for the type of merger
of compact objects depicted in Figure~4 of van den Bergh et al. Bright
inner  star--forming  regions  in  NGC  4258  (lower  right  panel  of
Figure~\ref{peculiar}) give the galaxy  a ``tadpole'' appearance at $z
\sim 2$,  similar to the  object in Figure~2  of van den Bergh  et al.
NGC 4258 could  also be interpreted as a multiple  merger on the basis
of  several   components  visible  in  the  image.    Along  with  the
irregular/peculiar   $A-C$   classifications   of   our   artificially
redshifted  galaxy  images,  the  individual examples  presented  here
suggest that one must approach with extreme caution the interpretation
of morphology in deep surveys.

\par  The  issue of  what  fraction  of  high--redshift galaxies  have
structure similar to local spirals rather than intrinsically irregular
morphologies bears  strongly on inferences of evolution  in the galaxy
population.  If the absence of ``ordered'' spirals at $z > 2$ noted by
Driver et  al. (1998)  is in part  due to bandshifting  effects, their
suggestion that the disk formation epoch occurs at $z \sim 2$ may need
to be modified.  Similarly, some of the compact objects at $z > 3$ may
be  the UV--bright  cores of  existing disk  galaxies rather  than the
early  progenitors   of  spheroidal  components   as  hypothesized  by
Giavalisco et al. (1996b).  The apparent peculiarity of high--redshift
galaxies has also prompted estimates of high merger rates at $z > 1.5$
(Driver et  al.) and an  increased merger fraction with  redshift (van
den Bergh  et al. 1996).  However,  merger rates at  high redshift are
particularly difficult to estimate due to the fragmented appearance in
the rest--frame UV of even undisturbed galaxies. As noted by Bunker et
al. (1999) and \cite{hib97}, it  is likely that imaging at longer rest
wavelengths  will be  necessary to  determine if  there is  a regular,
underlying  structure of  older  stars in  any  of the  high--redshift
galaxies.  Alternatively,  it would be  of interest to obtain  a large
sample of local  UV galaxy images such as that  proposed by the Galaxy
Evolution Explorer (GALEX, \cite{mar99}).   By comparing the local and
distant  UV  morphologies on  a  statistical  basis,  one could  infer
evolution in the UV properties  and thus study the history of physical
processes that  produce UV emission  without reference to  the optical
light.  Although  the full picture  of evolution would  necessarily be
incomplete without  data for the old  stars, an UV  database for local
galaxies would provide a direct comparison for distant systems.

\par  Our lack  of success  in classifying  FUV galaxies  in  terms of
Hubble types (in  this paper and K00) drives  us to consider alternate
methods of characterizing their morphology.  We therefore propose some
specific elements  of a classification  system that would  enhance our
descriptive power  for FUV images and  for irregular/peculiar galaxies
in general.  A  new method of defining galaxy  apertures that does not
assume a generally symmetric  distribution around a central brightness
enhancement,  would  greatly   extend  the  capabilities  to  describe
peculiar systems and mergers.  In  K00, we treat the merger system NGC
4038/9 (the  Antennae) as one galaxy  and define the aperture  to be a
circle enclosing  all pixels above $1.5\sigma_{\rm sky}$  that have at
least  four adjacent neighbors  also above  the threshold.   A similar
procedure could be used to define elliptical apertures for any galaxy.
It is also desirable to find parameters that are robust against errors
in centroiding: the  brightest part of the FUV image  is often not at,
or even near, the optical center, due to the prominence of off--center
star--forming  regions.   Instead  of  a {\it  central}  concentration
measurement, a  simple concentration index that  measures the fraction
of  total  light  in  a  small aperture  around  the  brightest  pixel
(wherever  it  is  located)  may  be more  appropriately  measured  on
irregular  galaxies.  Another  useful element  would be  the  scale of
various  features or  asymmetries.  This  would distinguish  between a
global disturbance  and a peculiarity  due to a small,  bright region.
One  could also  measure  the  size scale  of  the brightest  emitting
regions  to  determine if  most  of the  light  comes  from one  large
structure  or many  smaller  ones  and thus  describe  the texture  or
``patchiness'' of  the galaxy.  Finally,  it is important  to consider
the  effects  of  noise   on  any  derived  morphological  parameters,
especially for  rest--frame UV images  or faint galaxies.  In  our FUV
images, large regions within  the galaxy aperture often contain little
or  no  galaxy  light,  contributing   noise  but  no  signal  to  the
measurements  (particularly the  asymmetry).  \par  In the  future, we
intend to develop  quantitative morphological parameters applicable to
rest--frame UV  images, which could  serve as tools  for morphological
analysis of future UV imaging data from STIS and GALEX as well as deep
optical/NIR surveys of distant  objects.  A classification scheme with
the characteristics outlined above  could prove a valuable addition to
the Hubble sequence for studies  of evolution in the galaxy population
over the history of the universe.

\acknowledgements  The  authors  gratefully  acknowledge the  help  of
R. Bernstein, J.  Parker, R. Phelp s, and  N. Silbermann for obtaining
some of the optical imaging data presented here.  We thank O. Pevunova
for assistance  with preliminary data reduction.  Funding  for the UIT
project was provided through  the Spacelab office at NASA Headquarters
under Project Number 440-51.  WLF and BFM acknowledge support from the
Astro--2 Guest  Investigator Program through  grant number NAG8--1051.
Some  of the  data  presented here  were  obtained at  CTIO, which  is
operated by  AURA as part of  NOAO under a  cooperative agreement with
the National  Science Foundation.  This  research has made use  of the
NASA/IPAC Extragalactic  Database (NED), which is operated  by the Jet
Propulsion  Laboratory,  California  Institute  of  Technology,  under
contract with  NASA. We thank  the referee, Dr. Matthew  Bershady, for
helpful comments and his forebearance while this paper was revised.

Data discussed in this paper will be made publically available through
the NED (http://nedwww.ipac.caltech.edu) on an object-by-object basis,
and  as   a  collection  through  {\t  LEVEL5:   A  Knowledgebase  for
Extragalactic            Astronomy            and           Cosmology}
(http://nedwww.ipac.caltech.edu/level5).

\clearpage

\figcaption{FUV and artificially redshifted images of M101 and NGC 4214,
oriented with North up and East to the left.    The FUV M101 image is a
1311s UIT exposure, and the redshifted M101 images
are scaled as they would appear in the HDF.  The FUV NGC 4214
image is a 1061s UIT exposure, and the redshifted NGC 4214 images
again simulate the HDF.  The scale bar shown in the F300W panel for each
galaxy is applicable to all redshifted images of that galaxy. \label{fuvred}}
\figcaption{$U$--band and artificially redshifted images of M51, oriented 
with North up and East to the left.  The $U$--band image has
foreground stars removed.  The center and right--hand panels 
and simulate the appearance of M51 
in the HDF at the redshifts noted on the labels. \label{ured}}
\figcaption{Values of the central concentration ($C$) and asymmetry ($A$) 
indices as a function of wavelength for ten galaxies in our sample. 
\label{cawave}}
\figcaption{Change in $C$ and $A$ between optical and FUV images, for different
optical filters noted on the label in each panel.  Open circles are galaxies of 
type E/S0, solid squares are types Sa--Sb, open squares are Sbc--Sd, and stars
are types Irr/Pec/Merger.  Errorbars for $\Delta C$ and $\Delta A$ are
shown in the bottom panels. In the right--hand panels, solid 
horizontal lines mark $\Delta C,A = 0$, diagonal dotted lines are fits to data
from all filters together, and diagonal solid lines are fits to data for
each filter individually. \label{delca}}
\figcaption{Optical and FUV images of NGC 1512, NGC 3310, and NGC 4736 to 
demonstrate band--shifting effects in various galaxy types.  Images are
registered to the FUV coordinate system and oriented with North up and East to
the left.  As in K00, images are displayed in calibrated flux units such that pixels of the same surface brightness have constant darkness for both images
of the same galaxy. The scale bar in the lower right is applicable to all
images. \label{bandsh}}
\figcaption{Optical ($V$ or $R$ as available, $B$ for NGC 1068) and FUV galaxy
data on the log~$C$~$vs$~log~$A$ diagram.  Symbols for galaxies
of different Hubble types are the same as in Figure~\ref{delca}.
Dashed lines denote the divisions into morphological bins used by Abraham et al.(1996b), which have been shifted by the offsets calculated in Section~3.
Errorbars for different locations on
the diagram are shown as described in the text. \label{logca}}
\figcaption{Histograms of the change in $C$ and $A$ between the $U$--band or
FUV image and its artificially redshifted counterparts. The bin size is 0.1,
and errorbars for $\Delta C$ and $\Delta A$ based on the uncertainties in $C$
and $A$ (see text) are shown in the top panels.  Labels in each panel show the
rest--frame image used (either $U$ or FUV) and the redshift simulated.  
\label{delzca}}
\figcaption{Histograms of $\Delta C$ and $\Delta A$ resulting from changing
the aperture size from a fixed limit for all data to an adjusted value
based on the limiting surface brightness of each image.  Errorbars are shown
based on the propagation of our uncertainties in $C$ and $A$ through the 
arithmetic to get $\Delta C,A$.  This likely overestimates the true error, as
here we are comparing values of $C,A$ measured on the same image but
with different aperture radii.  \label{apcor}}
\figcaption{The apparent change in galaxy morphology with redshift in optical observations.  The left--hand panels show nearby galaxies as seen in the 
$V$--band, the center and right--hand panels show the same galaxies as they
would appear at the redshifts and in the WFPC2 filters noted on the labels.  The center panels are artificially redshifted rest--frame $U$--band images, and the
right--hand panels are rest--frame FUV images.   Each $V$--band image is scaled to the equivalent of a $\sim 5$ second exposure and binned by 3; the
artificially redshifted images are unbinned and represent $\sim 5\%$ of the HDF exposure time for the center panels and the entire HDF exposure time for the right--hand panels.  \label{3redgal}}  
\figcaption{The change in locus of galaxies on the log~$C$~$vs$~log~$A$ 
diagram with redshift.  Symbols for galaxies of different 
Hubble types are the same as in Figure~\ref{delca}, and dashed lines denote
the divisions into morphological bins as in Figure~\ref{logca}.  Representative
errorbars for different locations on the diagram are shown in the final panel,
as in Figure~\ref{logca}.  Data for $z \leq 1.2$ are from redshift simulations
of rest--frame $U$--band images, and data for $z \geq 2$ are from
artificially redshifted FUV images. \label{finlogca}}
\figcaption{Trajectories of several galaxies in the log~$C$~vs~log~$A$
diagram from their position as measured on optical images to that measured 
on artificially redshifted rest--frame UV images.
The larger point for each galaxy marks the optical position, then
smaller points at each simulated redshift are connected in order by the lines. 
\label{traj}}
\figcaption{Comparison to the HDF morphologies and redshift simulations of 
Abraham et al. (1996a).  The shaded region denotes the area occupied by HDF
galaxies, and the open polygon is the locus of the artificially redshifted 
Frei galaxy sample as measured by Abraham et al. (1996a).
Our data for different redshift simulations are shown as
different symbols for each redshift, as noted on the plot.  Errorbars for 
different locations on the diagram are shown. \label{hdfplot}}
\figcaption{Simulated high--redshift spirals that resemble peculiar galaxy types identified in deep survey images.
Images are oriented with North up and East to the left, and a scale bar 
and label for the simulated redshift are shown in each panel. \label{peculiar}}
\begin{table}
\dummytable\label{galxdat}
\end{table}
\begin{table}
\dummytable\label{cadat}
\end{table}
\begin{table}
\dummytable\label{redca}
\end{table}
\end{document}